\documentclass[a4paper,fleqn,usenatbib]{mnras}

\usepackage{mathptmx}

\usepackage[T1]{fontenc}
\usepackage{ae,aecompl}

\usepackage{graphicx}	
\usepackage{amsmath}	
\usepackage{amssymb}	
\usepackage{wasysym}           
\usepackage{epstopdf}
\usepackage{mathrsfs}
\usepackage{natbib}
\usepackage{color}
\usepackage{hyperref}

\title[Super-Eddington Accretion]{Super-Eddington Accretion in Tidal Disruption Events: the Impact of Realistic Fallback Rates on Accretion Rates}

\author[Wu, Coughlin \& Nixon]{Samantha Wu$^1$\thanks{email: samanthachloewu@berkeley.edu},
Eric R.~Coughlin$^1$\thanks{Einstein fellow}, Chris Nixon$^2$
\\
$^1$Astronomy Department and Theoretical Astrophysics Center, University of California, Berkeley, CA 94720, USA \\
$^2$Theoretical Astrophysics Group, Department of Physics \& Astronomy, University of Leicester, Leicester LE1 7RH UK
}

\date{Accepted XXX. Received YYY; in original form ZZZ}

\pubyear{2017}

\begin{document}
\label{firstpage}
\pagerange{\pageref{firstpage}--\pageref{lastpage}}
\maketitle

\begin{abstract}
After the tidal disruption of a star by a massive black hole, disrupted stellar debris can fall back to the hole at a rate significantly exceeding its Eddington limit. To understand how black hole mass affects the duration of super-Eddington accretion in tidal disruption events, we first run a suite of simulations of the disruption of a Solar-like star by a supermassive black hole of varying mass to directly measure the fallback rate onto the hole, and we compare these fallback rates to the analytic predictions of the ``frozen-in'' model. Then, adopting a Zero-Bernoulli Accretion flow as an analytic prescription for the accretion flow around the hole, we investigate how the accretion rate onto the black hole evolves with the more accurate fallback rates calculated from the simulations. We find that numerically-simulated fallback rates yield accretion rates onto the hole that can, depending on the black hole mass, be nearly an order of magnitude larger than those predicted by the frozen-in approximation. Our results place new limits on the maximum black hole mass for which super-Eddington accretion occurs in tidal disruption events. 
\end{abstract}

%
\section{Introduction}

When stars pass within a distance $r_{\rm t} \simeq (M_{\rm h}/M_{*})^{1/3}R_*$ of a supermassive black hole (SMBH), where $M_h$ is the hole mass and $M_*$ and $R_*$ are respectively the mass and radius of the star, they are disrupted by the tidal field of the SMBH. Upon disruption, half of the stellar mass is ejected from the system while the other half remains bound to the black hole on initially-Keplerian orbits (e.g., \citealt{rees88,evans89}). This bound material falls back to the SMBH and forms an accretion disk on a timescale that is likely comparable to the period of one complete orbit of the most bound material \citep{bonnerot16b,hayasaki16,shiokawa15,sadowski16}, also called the fallback time, and generates quasar-like emission for timescales of months to years. Throughout the process, the fallback rate is predicted to follow a $t^{-5/3}$ decline \citep{phinney89}, though deviations from this canonical power-law can arise for a variety of reasons (e.g., \citealt{lodato09,lodato11,guillochon13,hayasaki13,coughlin15,coughlin17}). This sequence of events, from the initial disruption of the star to the fallback of tidally disrupted debris and formation of an accretion disk, is known as a tidal disruption event (TDE). 

In some TDEs, the fallback rate can exceed the Eddington luminosity of the SMBH, making winds \citep{strubbe11} and jets \citep{giannios11} a likely byproduct. Observational evidence accumulated over the past three decades has revealed TDE candidates (see \citealt{komossa15} for a review), and more recently, jetted TDEs have been discovered that are likely associated with super-Eddington accretion. For example, \emph{Swift} J1644+57 \citep{burrows11,bloom11,levan11,zauderer11} is now interpreted as the first known example of a jetted TDE and reached peak luminosities in excess of $10^{47}$ erg s$^{-1}$ - above the Eddington limit for nearly every model of the SMBH powering the event. Since then, two other candidates for jetted, likely super-Eddington TDEs have been detected \citep{cenko12,brown15}, and others, such as ASASSN-14li \citep{miller15,alexander16,alexander17}, that show wind-like emission have also been found (though there is also evidence for a faster, jetted outflow at earlier times in this system; \citealt{vanvelzen16,pasham17,kara18}). 



From a theoretical standpoint, accurately simulating a TDE with realistic parameters (e.g., a Solar-like progenitor and a $10^6$ M$_{\odot}$ SMBH) is extremely computationally expensive. A simulation that captures the TDE from the initial stellar disruption to the formation and evolution of the accretion disk must resolve not only the stellar radius ($\sim 1R_{\odot}$), but also the thousands of stellar radii to which the debris stream expands. To additionally model the disk that forms from the debris, one must, at the very least, account for general relativistic precession, which is thought to be the dominant mechanism responsible for dissipating angular momentum and facilitating the circularisation of the debris \citep{rees88}, and radiation feedback when the accretion rate onto the SMBH is super-Eddington.

Simulations investigating the disk formation following a TDE (e.g., \citealt{guillochon14,bonnerot16b,hayasaki16,shiokawa15}) have shown that the debris disk extends to large radii much greater than the tidal radius and is geometrically thick when the radiation is trapped. Consequently, the material cannot cool; this condition is especially true when the accretion rate onto the SMBH is super-Eddington \citep{begelman79}. \citet{sadowski16} additionally found that, in their simulation of the disk formation following the disruption of a $0.1M_{\odot}$ star by a $10^5M_{\odot}$ SMBH, the debris conformed to a configuration well-described by a Bernoulli function, or specific energy, that was globally equal to approximately zero. In their simulation, the peak accretion rate onto the SMBH was in excess of $\sim 10^4$ L$_{\rm Edd}$.

\citet{coughlin14}, hereafter CB14, proposed that a zero-Bernoulli accretion (ZEBRA) flow - one for which the specific energy is globally equal to zero - should describe the disk formed during the super-Eddington phase of a TDE. In particular, they argued that the highly super-Eddington accretion and inefficient cooling causes the debris disk to approach a weakly-bound (zero-Bernoulli), highly-inflated state. Upon reaching this zero-Bernoulli configuration, the remaining accretion energy is freed as material falls onto the SMBH at the disk center. This accretion energy is then anisotropically funneled into two bipolar jets that escape along the rotational axis of the flow, thereby safely removing the accretion energy from the system without destroying the accretion structure in the process. The disk evolution is then regulated by the mass fallback rate, the black hole accretion rate, and the total angular momentum of the flow, and the disk properties (e.g, its angular momentum and density profiles) conform to simple, analytic functions. This prescription for the accretion disk structure is  actualized in the numerical simulations of \citet{sadowski16} and qualitatively consistent with observed jetted TDEs such as \emph{Swift} J1644+57. 

In this paper, our goal is to assess the duration and magnitude of super-Eddington accretion in TDEs as a function of black hole mass. We first simulate the tidal disruption of a Solar-like star by a black hole of varying mass and directly measure the fallback rate. Then, instead of taking the most rigorous, but prohibitively computationally expensive, approach of numerically simulating the disk formation and evolution, we use the ZEBRA model of CB14 to determine the structure of the accretion disk and its accretion rate as a function of the fallback rate. In this way we obtain a very accurate measure of the rate at which the ZEBRA flow is fed, for which past investigations (e.g., CB14) have only used approximate, analytic formulas, and our prescription for the disk structure allows us to follow the long-term evolution of the disk during the super-Eddington phase. This method allows us to place tight constraints on the timescale over which TDEs maintain super-Eddington accretion and the maximum mass capable of powering super-Eddington TDEs.  

In Section \ref{sec:simulations} we present the simulations of the disruption of the star and the fallback rates obtained therefrom. In Section \ref{sec:accretion} we outline the zero-Bernoulli Accretion model for the accretion flow during the super-Eddington accretion phase of the TDE, and we present the time-dependent evolution of the disk and luminosity. We discuss the implications and interpretation of our findings in Section \ref{sec:discussion}, and we summarize and conclude in Section \ref{sec:summary}.

\section{Simulations}
\label{sec:simulations}
We used the Smoothed-particle hydrodynamics (SPH) code {\sc phantom} \citep{price17} to simulate the tidal disruption of a Solar-like star (i.e., stars with a Solar mass and radius) by a supermassive black hole of variable mass. The star is modeled as a polytrope (e.g., \citealt{hansen04}) with polytropic index $\gamma = 5/3$, which provides a reasonable approximation for the density profile of the Sun and less massive stars. The polytropic profile is achieved by placing $10^7$ particles on a close-packed sphere that is subsequently stretched to approximate the density distribution; numerically-induced perturbations  are then smoothed by relaxing the star for ten sound crossing times in isolation. 

The relaxed polytrope is then placed at 5 $r_{\rm t}$ from the hole on a parabolic orbit with $\beta \equiv r_{\rm t}/r_{\rm p} = 1$, where $r_{\rm t} = R_*\left(M_{\rm h}/M_*\right)^{1/3}$ is the tidal radius and $r_{\rm p}$ is the pericenter distance of the stellar center of mass from the SMBH. The SMBH is modeled as a Newtonian point mass with an ``accretion radius,'' such that any particle entering within that radius is removed from the simulation. We include the effects of self-gravity through the usage of a bisective tree algorithm alongside an opening angle criterion \citep{gafton11}. The gas is assumed to evolve adiabatically with adiabatic index $\gamma = 5/3$, which provides a good approximation for the early evolution of the disrupted stellar debris (\citealt{coughlin16}, though the effects of magnetic fields and radiative recombinations can invalidate the isentropic assumption at late times; \citealt{guillochon17, bonnerot17,kasen10}).

\begin{figure*}
\includegraphics[width=0.495\textwidth]{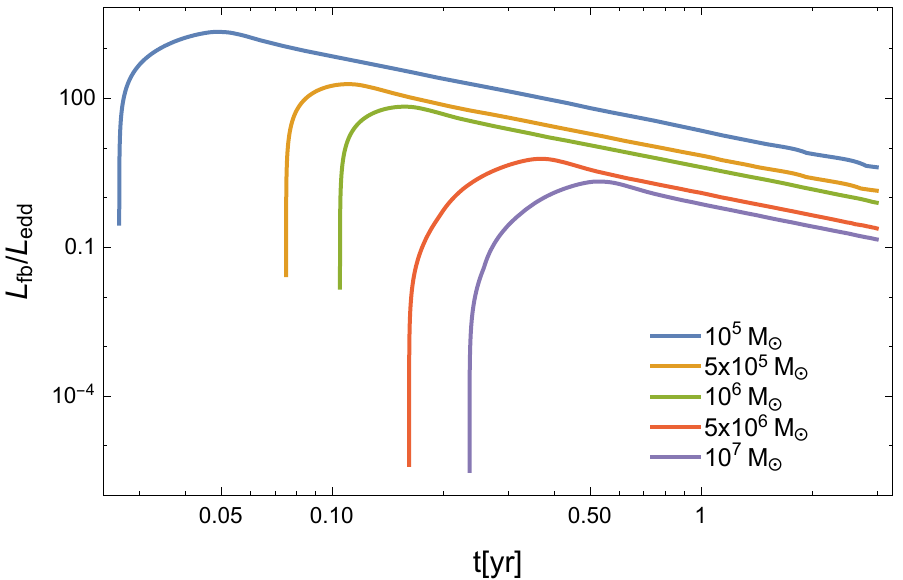}
\includegraphics[width=0.495\textwidth]{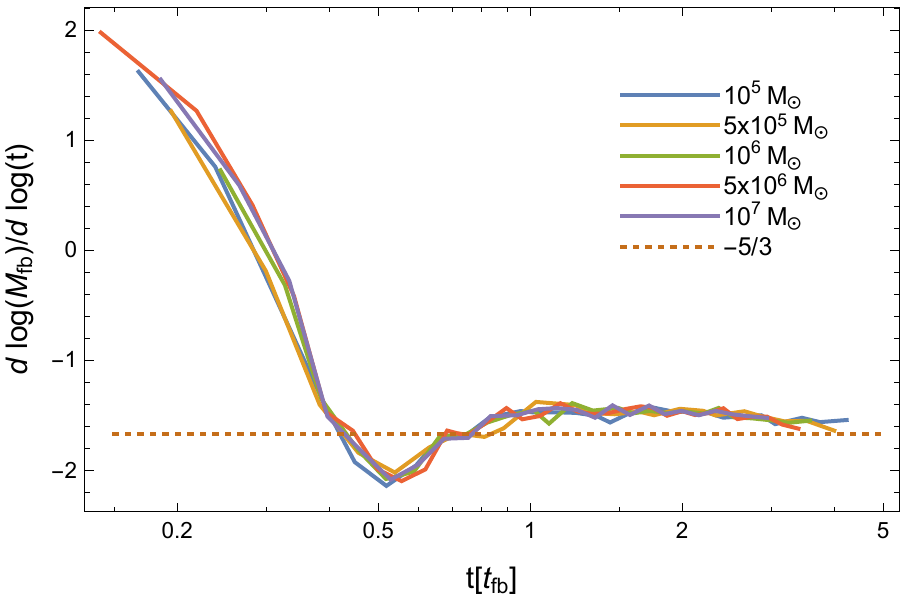}
\caption{Left Panel: The fallback luminosity $L_{\rm fb}$ resulting from the SPH simulations for the black holes listed in the legend, calculated  assuming a radiative efficiency of $\epsilon = 0.1$ (i.e. $L_{\rm fb} = \epsilon \times \dot{M}_{\rm fb}c^2$, where $\dot{M}_{\rm fb}$ is the fallback rate). In each case we normalized the fallback luminosity to the Eddington limit of the hole $L_{\rm Edd} = 4\pi GM_{\rm h}c/\kappa$, where $\kappa = 0.34 $ cm$^2$ g$^{-1}$ is the Thomson opacity assuming standard abundances. Right Panel: Logarithmic derivative of the numerical fallback rate as a function of time, or the instantaneous power-law slope of the fallback rate, in units of the fallback time for each black hole mass. The logarithmic derivative shows the power law index of $\dot{M}_{\rm fb}$ for each black hole mass, which in the figure approaches but remains above the theoretical value of $t^{-5/3}$ at late times for each black hole mass.}
\label{fig:dmdtplots}
\end{figure*}
\begin{figure}
\includegraphics[width=0.495\textwidth]{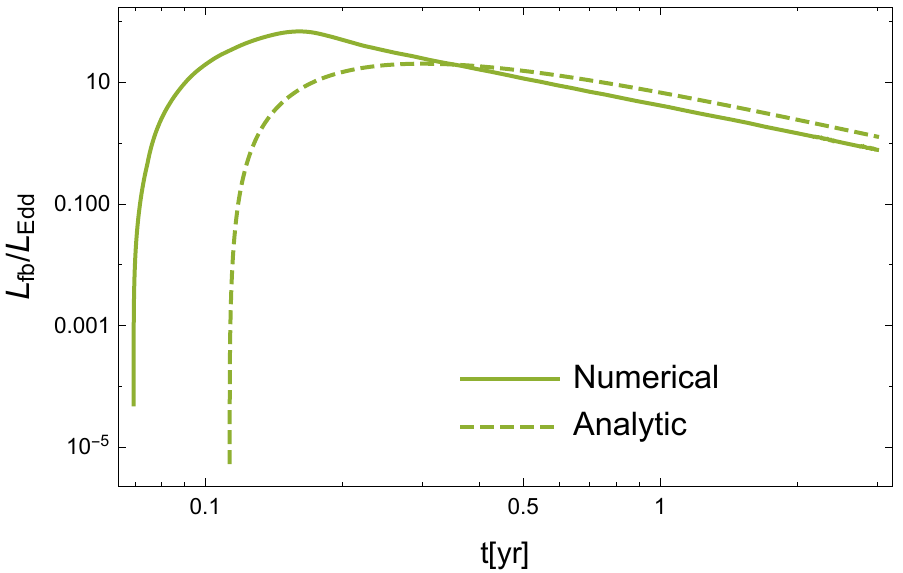}
\caption{The fallback luminosity $L_{\rm fb}$ for $10^6M_{\odot}$, again assuming a radiative efficiency of $\epsilon = 0.1$ and normalized to the Eddington limit of the hole. The numerical $L_{\rm fb}$ resulting from the ZEBRA model with ${\dot m}_{\rm fb}(t)$ from the SPH simulations is shown in solid green, while the analytic $L_{\rm fb}$ is shown dashed.}
\label{fig:dmdt16}
\end{figure}
\begin{figure*}
\includegraphics[width=0.495\textwidth]{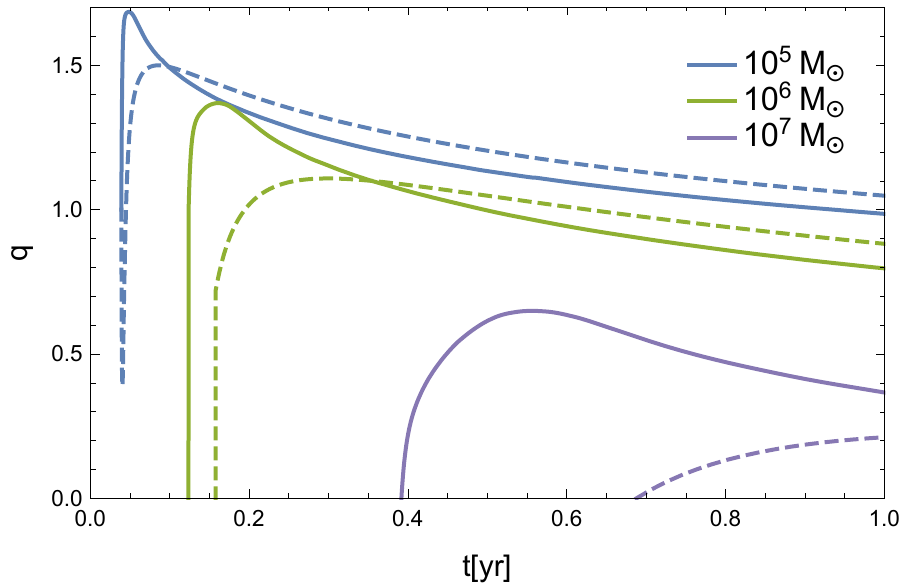}
\includegraphics[width=0.495\textwidth]{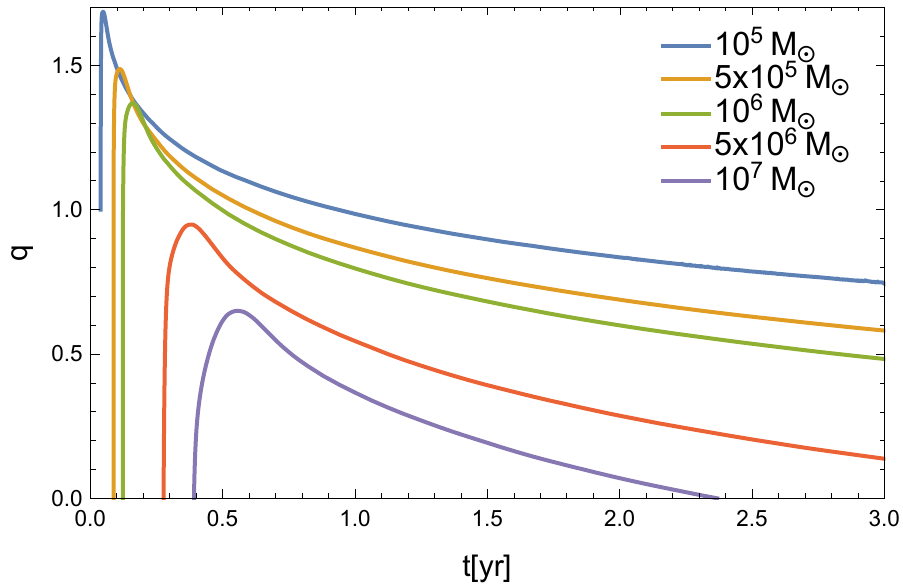}
\caption{Left Panel: The parameter $q(t)$ (defined in Equation \eqref{rho}) that determines how quickly the density falls off in spherical radius $r$, for the black hole masses indicated in the legend. The results of using the numerically-acquired fallback rate are plotted in solid lines, whereas the solution from the frozen-in approximation is shown dashed. Right Panel: Numerical results for $q(t)$ for the indicated black hole masses.}
\label{fig:qplots}
\end{figure*}

During the initial disruption, the accretion radius of the SMBH is set well within the tidal radius. Since we are interested in the rate of return of bound material, the fallback rate $\dot{M}_{\rm fb}$, to the SMBH, we set the accretion radius to 3 r$_{\rm t}$ once the disrupted debris stream is well beyond the tidal radius but before the most bound segment of the stream has returned to pericenter. Particles that return to the SMBH are then ``accreted'' and removed from the simulation upon passing through that radius, and the total number of particles accreted per unit time defines the fallback rate. We therefore do not simulate the initial formation of the disk, which would (at the very least) necessitate a much greater particle number, an inclusion of post-Newtonian terms, and an incorporation of radiation-hydrodynamics to accurately capture the physics of recompression shocks, relativistic precession, and super-Eddington feedback. That the disk is not initially in a quasi-steady state, and instead forms chaotically as a result of these effects, is substantiated by the early, erratic behavior of Swift J1644, which could be indicative of hysteresis regarding the jet direction \citep{tchekhovskoy14}. However, if the disk is formed promptly through the combination of these effects, then the fallback rate we measure from the simulation is equivalent to the rate at which material is incorporated into the accretion flow. We will assume here that this scenario is actualized, though the situation could be more complicated if, for example, nodal precession delays stream-stream intersections \citep{guillochon15} or the stream is adversely influenced by the presence of a nascent ambient medium \citep{bonnerot16b,kathirgamaraju17}.  

\begin{figure}
\includegraphics[width=0.5\textwidth]{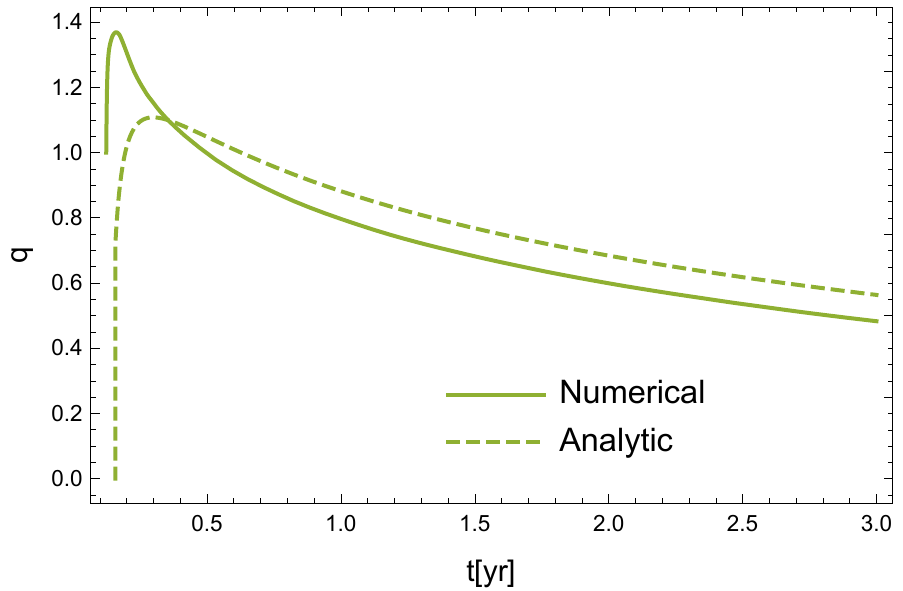}
\caption{As in Figure \ref{fig:qplots}, the parameter $q(t)$ for a black hole with mass $10^6M_{\odot}$. Numerical results are plotted in solid green and analytic results are shown dashed.}
\label{fig:q16}
\end{figure}

We simulated the disruption of a Solar-like star by a SMBH of mass $10^5$, $5\times 10^5$, $10^6$, $5\times 10^6$, and $10^7$ $M_{\odot}$, and measured the resulting fallback rates. Figure \ref{fig:dmdtplots}, left panel, shows the numerically-determined fallback rates normalized by the Eddington luminosity $L_{\rm Edd} = 4\pi GM_hc/\kappa$ of the hole, where the radiative efficiency $\epsilon$ we set to 0.1 and $\kappa = 0.34$ cm$^2$ g$^{-1}$ is the Thomson opacity, for a range of black hole masses. The fallback rate as a function of time for each black hole mass approaches a power law at later times. 

The right-hand panel of Figure \ref{fig:dmdtplots} gives the logarithmic derivative of the log of the fallback rate, which traces the instantaneous power-law index of the fallback curves in the left-hand panel of the same figure. Here time is measured in units of the fallback time of the appropriate black hole, and hence the theoretically-predicted dependence of this result on the black hole mass ($\propto M_{\rm h}^{1/2}$) has been scaled out. The fall below and rise above the asymptotic $-5/3$ limit is in agreement with the simulations of \citet{guillochon13}, and the fact that this oscillatory behavior is almost completely independent of the black hole mass points to its physical origin as related to the internal properties of the gas. In particular, it is likely self-gravity, which draws material from the radial extremities of the stream and induces density perturbations on top of the analytically-predicted profile, that acts on the sound crossing time (which is much less than the fallback time for the SMBHs considered here) to generate these features. This notion is supported by the fact that streams with softer equations of state, which are less prone to the effects of self-gravity \citep{coughlin16}, exhibit less deviation in their fallback rates \citep{coughlin16b}.

Figure \ref{fig:dmdt16} depicts the fallback luminosity $L_{\rm fb} = \epsilon \dot{M}_{\rm fb} c^2$, where we have assumed a radiative efficiency of $\epsilon = 0.1$, for both analytic and numerical prescriptions for the fallback rate $\dot{M}_{\rm fb}$. 
The analytic fallback rate shown in this figure is derived from the ``frozen-in,'' or impulse approximation (\citealt{lodato09}; see also CB14, who write the explicit expression for the fallback rate for polytropes in their Equation 34). In this approximation, we assume the entire star remains in hydrostatic equilibrium and moves with its center of mass -- which follows a parabolic orbit -- until reaching the tidal radius. At this point the tidal force of the SMBH overwhelms the self-gravity and pressure of the star, and fluid parcels thereafter follow ballistic orbits in the gravitational field. Therefore, this approximation treats the tidal force as acting impulsively at the tidal radius, and the Keplerian energies of the gas parcels comprising the star are frozen in at that time. 

It is evident from Figure \ref{fig:dmdt16} that our numerical, more accurate approach yields a higher peak fallback rate (by nearly an order of magnitude) and an earlier return time of the most bound debris than the analytic prescription (e.g., \citealt{lodato09, coughlin15}). 
Since the fallback of stellar debris feeds  the accretion disk, we expect the higher fallback rate to correspond to a higher accretion rate onto the SMBH. In the next section, we use these fallback rates with the analytic ZEBRA model of CB14 to investigate the accretion rate onto the SMBH and the time-dependent structure of the accretion disk. 



\section{Accretion}
\label{sec:accretion}
\subsection{Accretion disk structure}

From Figure \ref{fig:dmdtplots}, we see that the early stages of a TDE by SMBHs with mass $\lesssim 10^{7}M_{\odot}$ are super-Eddington. During this phase, if shocks are efficient at circularizing the debris, the extreme luminosity of the black hole increases the material's specific energy to the point where its Bernoulli parameter nears zero, resulting in a very weakly-bound disk. 

Once the gas reaches the zero-Bernoulli condition, the momentum equations permit self-similar solutions showing that the marginally-bound disk forms a quasi-spherical envelope that is closed except at the poles, and CB14 coined this a Zero-Bernoulli accretion, or ZEBRA, flow. The closed nature of the disk leaves no surface from which the energy released during the super-Eddington accretion may be exhausted (e.g., through winds). As a result, CB14 postulated that the accretion energy in the disk would be directed to the poles, launching jets to exhaust the energy. 



While it is possible to construct very general solutions for the density, angular momentum, and pressure profiles of a ZEBRA envelope (see Appendix A of CB14), it is likely that these quantities vary approximately self-similarly from an inner radius $r_0$, which is near the inner most stable circular orbit of the gas, out to the trapping radius \citep{begelman79} (outside of which photon diffusion becomes efficient and the gas can cool and transition to a thin disk). In the self-similar limit, the density ($\rho$), pressure ($p$), and square of the specific angular momentum ($\ell^2$) of a ZEBRA envelope, which result from the radial and polar momentum equations alongside the zero-Bernoulli criterion, are

\begin{equation}
\rho(r,\theta) = \rho_0\left(\frac{r}{r_0}\right)^{-q}(\sin^2{\theta})^{\alpha} \label{rho}
\end{equation}
\begin{equation}
p(r,\theta)=\beta\frac{GM_h}{r}\rho \label{peq}
\end{equation}
\begin{equation}
\ell^2(r,\theta)=aGM_hr\sin^2{\theta} \label{ell}
\end{equation}
where the constants $\alpha$, $\beta$, and $a$ satisfy

\begin{equation}
\alpha=\frac{1-q(\gamma-1)}{\gamma-1},\quad\beta=\frac{\gamma-1}{1+\gamma -q(\gamma-1)},\quad a=2\alpha\beta, \label{alpha}
\end{equation}
where $\gamma$ is the adiabatic index of the gas. Since super-Eddington accretion flows are dominated by radiation pressure, we will henceforth adopt $\gamma = 4/3$. Note that the $\beta$ appearing in Equations \eqref{peq} and \eqref{alpha} is \emph{not} the standard $\beta$ used in the TDE literature.
\\

We see from Equations \eqref{rho} -- \eqref{alpha} that a ZEBRA flow is characterized entirely by the number $q$, with larger (smaller) values corresponding to more (less) spherically-symmetric flow. 
Note from the expression for $\alpha$ that $q = 3$ gives $\alpha = 0$ when $\gamma = 4/3$, which is the correct solution for spherically-symmetric, radiation-dominated flow about a point mass. In their ADIOS models of accretion and outflow, \citet{blandford99,blandford04} related this parameter to the vigorousness of mass loss due to disk winds (which are ultimately driven by radiation pressure and magnetic fields), and it can therefore vary from system to system. However, in a tidal disruption event, the total angular momentum and mass contained in the disk are well-constrained by the properties of the disrupted star and the SMBH, and this suffices to uniquely determine $q$. Indeed, as shown in CB14, $q$ is related to the envelope mass $\mathscr{M}$, the envelope angular momentum $\mathscr{L}$, and the black hole mass $M_{\rm h}$ via the implicit equation

\begin{equation}
\frac{\Gamma(\alpha+1)^{5/6}\Gamma(\alpha+2)^{5/6}}{\beta^{1/6}a^{1/2}\Gamma(\alpha+3/2)^{5/3}}\frac{(7/2-q)^{5/6}}{3-q} = \left(\frac{y\kappa}{4\pi c}\right)^{1/6}\frac{\mathcal{M}\sqrt{GM_h}}{\mathcal{L}^{5/6}}, \label{qeq}
\end{equation}  
where $\Gamma$ is the generalized factorial and $y$ is a number of order unity (note that $y$ does not significantly impact the solution for $q$, as it only enters into the above equation to the 1/6 power). 


On physical grounds, $q$ must satisfy $q>0.5$ to ensure that energy generation in the disk decreases outwards. Similarly, $q$ must be less than its spherically-symmetric value ($q=3$ for a radiation-pressure dominated gas) to maintain a finite density at the poles. The permissible range of $q$ is therefore bounded by $0.5 < q < 3$. 




From Equation \eqref{qeq}, we find that the parameter $q$ varies monotonically with the ratio of the mass of the disk to its angular momentum, ranging from $0.5$ (small ratio of mass to angular momentum) to its spherically-symmetric value of 3 (large ratio of mass to angular momentum); see Figure 1 in CB14. In a TDE, we therefore expect $q$ to decrease over time as accretion onto the black hole depletes the mass contained in the envelope and angular momentum is transported outward. Equations \eqref{rho} -- \eqref{qeq} thus hold approximately at an instant in time\footnote{It should be noted that the self-similar solutions for the ZEBRA envelope ignore the explicit time dependence of the disk quantities in the momentum equations. However, these corrections should be small -- of order the ratio of the ZEBRA sound crossing time to the fallback time of the debris.}, and we can use the fallback rates numerically obtained in Section \ref{sec:simulations} to determine the time-dependent evolution of the disk properties. 

\subsection{Accretion rates and time-dependent evolution}

\begin{figure*}
\includegraphics[width=0.495\textwidth]{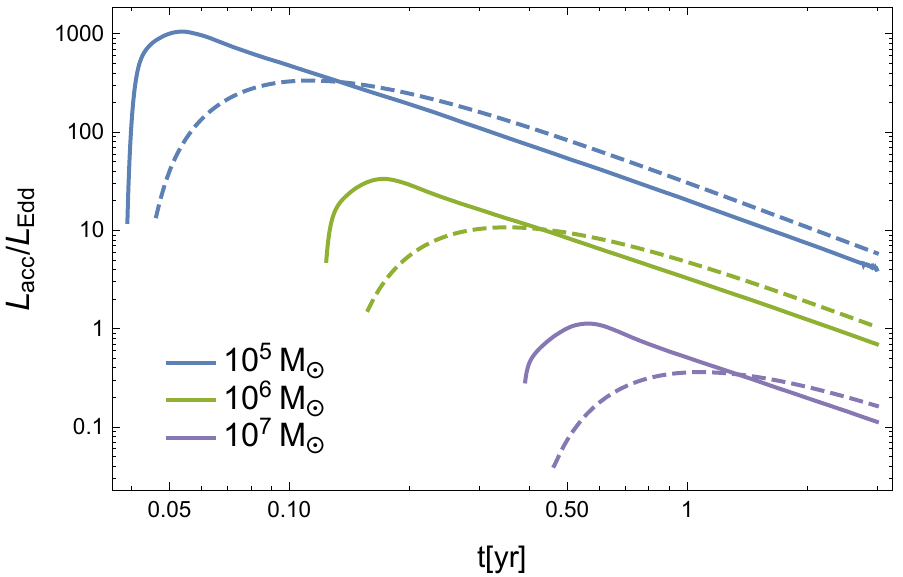}
\includegraphics[width=0.495\textwidth]{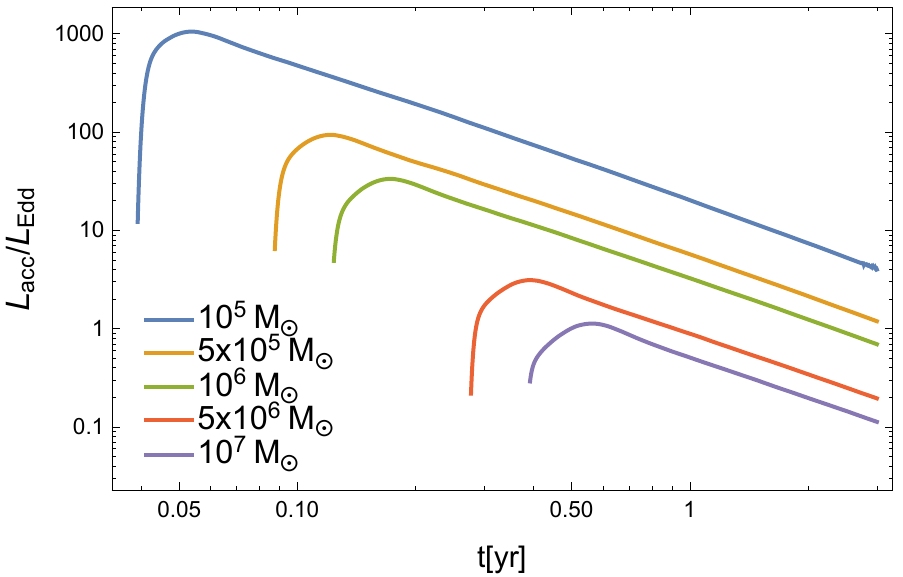}
\caption{Left Panel: The accretion luminosity $L_{\rm acc}$ resulting applying the ZEBRA model with ${\dot M}_{\rm fb}(t)$ from the SPH simulations (solid lines) and from the frozen-in approximation (dotted) for the black holes listed in the legend, assuming a radiative efficiency of $\epsilon = 0.1$ (i.e. $L_{\rm acc} = \epsilon \times \dot{M}_{\rm acc}c^2$). In each case we normalized the accretion luminosity to the Eddington limit of the hole. Right Panel: Accretion luminosity $L_{acc}$ calculated using ${\dot M}_{\rm fb}(t)$ from the SPH simulations normalized by the Eddington luminosity for the SMBH masses listed in the legend.}
\label{fig:maccplots}
\end{figure*}

\begin{figure*}
\includegraphics[width=0.495\textwidth]{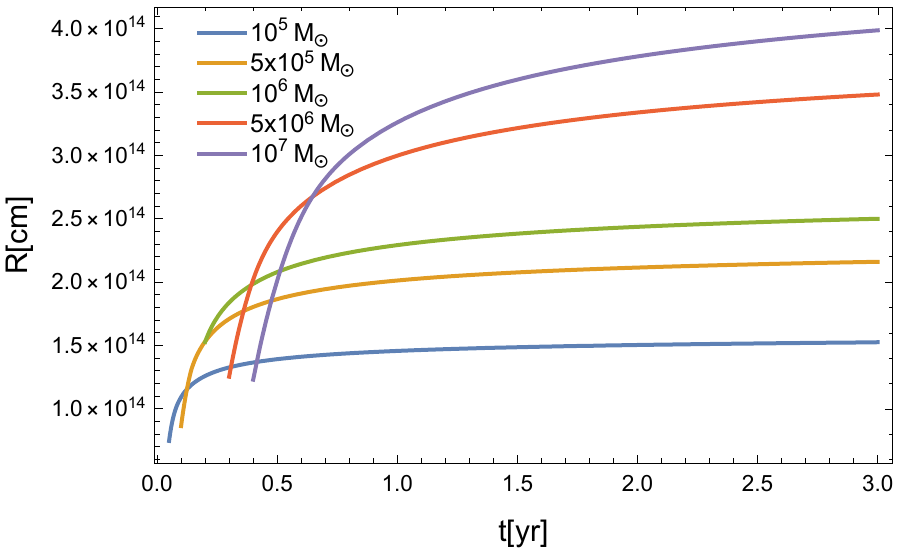}
\includegraphics[width=0.495\textwidth]{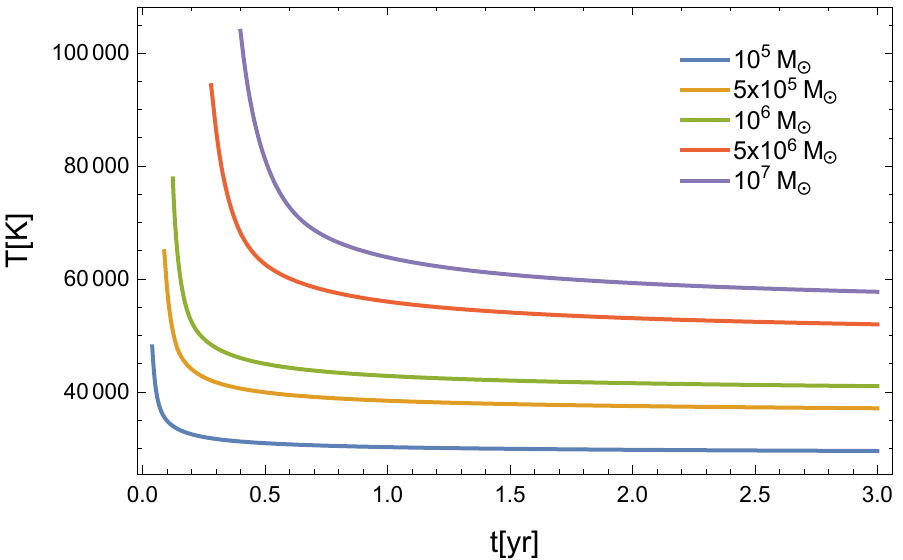}
\caption{Left Panel: The photospheric radius of the ZEBRA as a function of time, calculated using the numerical fallback rate for the black hole masses listed in the legend. Right Panel: Effective temperature of the photosphere as a function of time, also calculated using the numerical fallback rate for the black hole masses listed in the legend.}
\label{fig:rTplots}
\end{figure*}

The time-dependent evolution of the disk is governed by the global conservation of mass and angular momentum. The former is encapsulated by the differential equation $\dot{\mathscr{M}} = \dot{M}_{\rm fb}-\dot{M}_{\rm acc}$, where $\dot{M}_{\rm fb}$ is the fallback rate that feeds the disk and $\dot{M}_{\rm acc}$ is the accretion rate onto the black hole. $\dot{M}_{\rm fb}$ is determined numerically in Section \ref{sec:simulations}, and $\dot{M}_{\rm acc}$ can be found from the ZEBRA model and the assumption that the gas extends approximately to the innermost stable circular orbit (see Equations 26 and 27 of CB14). Since the angular momentum of the material must be lost as it falls onto the SMBH, conservation of angular momentum implies that $\mathscr{L} = \mathcal{L}_{\rm fb}$, where $\mathcal{L}_{\rm fb}$ is the angular momentum of the returning debris and is related to $\dot{M}_{\rm fb}$ and the angular momentum of the star at the time of disruption (see Equation 37 of CB14)\footnote{This assumption ignores the angular momentum lost through the ISCO, which should be small compared to the angular momentum added by the infalling debris}. These differential equations, coupled to the algebraic equation relating $q$, $\mathscr{M}$, and $\mathscr{L}$ (Equation \ref{qeq}), then permit a solution for $q(t)$ (and other time-dependent properties of the disk).  

To solve the differential-algebraic equations numerically, the initial value of $q$ must be specified. While this initial value is in principle determined from the efficiency of circularization and the configuration of the debris at the return time of the most bound debris, CB14 showed that solutions with different initial $q$ quickly converge to a unique solution after displaying transient behavior at very early times. As a result, the initial value of $q$ can be chosen arbitrarily within the permissible range $0.5 < q <3$, subject to the caveat that the initial, erratic behavior of the solutions is not adequately captured by our model. 


Numerical integration gives $q(t)$, as depicted for different black hole masses in the right-hand panel of Figure \ref{fig:qplots}. This is compared to the solution for $q(t)$ calculated using the frozen-in approximation, which gives an analytic prescription for the properties of the accretion disk as described in Section \ref{sec:simulations}. The left-hand panel of Figure \ref{fig:qplots} shows how the behavior of $q$ differs due to integration of the differential equation with the fallback rate acquired numerically as opposed to the frozen-in approximation for $\dot{M}_{\rm fb}$. For an interval on the order of 1 year after the return time of the most bound debris, $q(t)$ is higher from the simulation method than the analytic results, but after this period the results of using the simulated $\dot{M}_{\rm fb}$ decrease more rapidly from the peak and fall below the analytic prediction. Figure \ref{fig:q16} exemplifies this for $10^6 M_{\odot}$, showing that the peak for the numerical results reaches $q=1.4$, whereas the results for $q(t)$ from the frozen-in approximation only attain a peak of $q=1.1$.

Given the solution for $q(t)$, we compute the accretion rate onto the black hole, $\dot{M}_{\rm acc}$ using Equation 26 of CB14. The accretion luminosity $L_{acc}$ follows as $L_{\rm acc} = \epsilon\dot{M_{\rm acc}}c^2$, with $\epsilon=0.1$. The right-hand panel of Figure \ref{fig:maccplots} shows the accretion luminosity derived from the simulated fallback rate for different black hole masses, each normalized by its Eddington luminosity $L_{\rm Edd} = 4\pi GM_hc/\kappa$. Here $\kappa = 0.34 $ cm$^2$ g$^{-1}$ is the Thomson opacity assuming standard abundances. The simulated fallback rate leads to a higher accretion luminosity at earlier times, which then drops off more steeply and falls below the predictions of the frozen-in approximation within several months, as depicted in the left-hand panel of Figure \ref{fig:maccplots}. 

\begin{figure}
\includegraphics[width=0.495\textwidth]{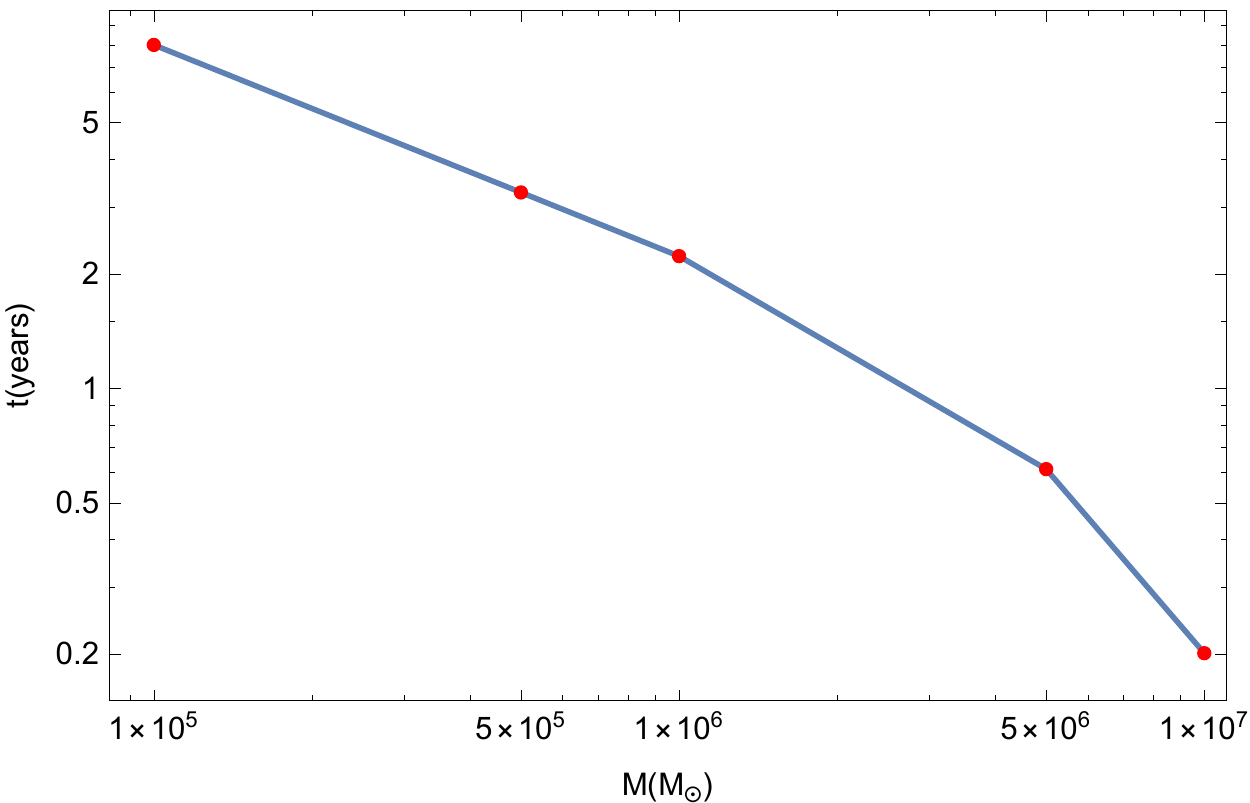}
\caption{Time at which the accretion luminosity of the ZEBRA disk is no longer super-Eddington, shown as a function of $M_{\rm h}/M_{\odot}$. Points in red denote the values for the black hole masses simulated in this paper, while the blue line is not a best fit but merely for clarity of visualization.}
\label{fig:supereddtime}
\end{figure}

\begin{figure}
\includegraphics[width=0.495\textwidth]{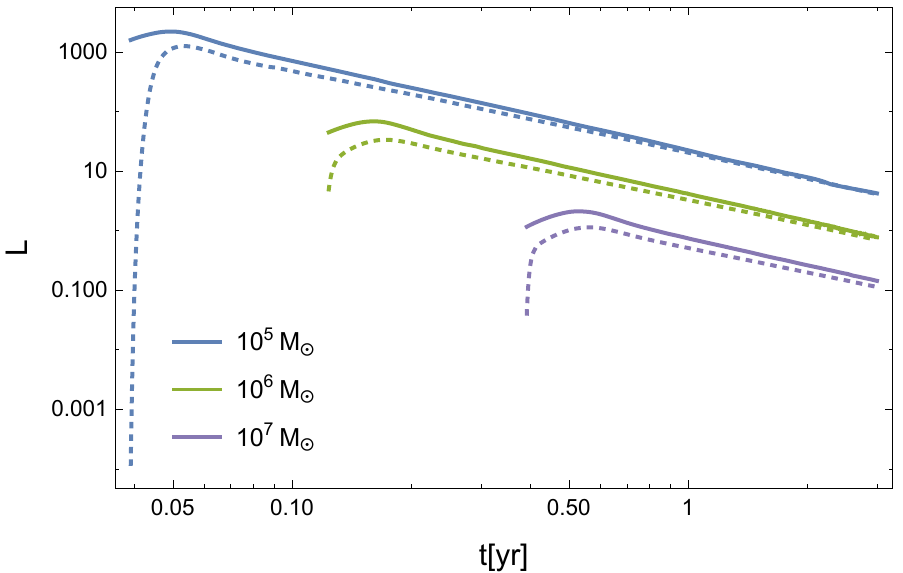}
\caption{The accretion luminosity $L_{\rm acc}$ resulting from the ZEBRA model with ${\dot M}_{\rm fb}(t)$  from the SPH simulations (dotted) versus the fallback luminosity  $L_{\rm fb} = \epsilon \times {\dot M}_{\rm fb}c^2$ (solid) for the black holes listed in the legend. In each case we assumed a radiative efficiency of $\epsilon = 0.1$ and normalized the luminosity to the Eddington limit of the black hole.}
\label{fig:maccvsfb}
\end{figure}

Furthermore, the 
rate at which material accretes onto the black hole is not exactly equal to rate at which debris falls onto the disk. Figure \ref{fig:maccvsfb} shows that the accretion luminosity follows a slightly steeper power law than the fallback luminosity, where both are calculated from simulation results for the fallback rate $\dot{M}_{\rm fb}$. This behavior is comparable to that shown in CB14, in which the frozen-in approximation yields a fallback rate proportional to $t^{-5/3}$ and an accretion rate that follows a slightly steeper power law (see equation 34 of CB14). The fallback rate also attains higher maximum values than the accretion rate, as in Figure \ref{fig:maccvsfb}.

We also determine the radius of the photosphere using the numerical values for ${\dot M}_{\rm fb}$ and ${\dot M}_{\rm acc}$. From CB14, the radius $\mathcal{R}$ is given by

\begin{equation}
\mathcal{R} = \left(\frac{y\kappa \beta \sqrt[]{a}(3-q)}{4 \pi c}\mathcal{M}\sqrt{G M_{\rm h}}  \right)^{2/5},
\end{equation}
where $\mathcal{M}$ is the total mass contained in the disc. We find $\mathcal{M}$ by numerically integrating the differential equation $\dot{\mathcal{M}} = {\dot M}_{\rm fb} - {\dot M}_{\rm acc}$. Since the photospheric radius coincides with the trapping radius \citep{begelman78}, which is where photon diffusion becomes efficient, the effective temperature at the photosphere is 

\begin{equation} 
T= \left(\frac{L_{\rm Edd}}{4\pi\sigma_{SB}\mathcal{R}^2}\right)^{1/4},
\end{equation} 
where $\sigma_{\rm SB} = 5.67\times 10^{-5}$ [cgs] is the Stefan-Boltzmann constant. The resulting photospheric radii and effective temperatures as functions of time are depicted in Figure \ref{fig:rTplots}.

From Figure \ref{fig:maccplots}, we see that the accretion luminosity remains super-Eddington for finite periods that differ for each black hole mass. The time at which the transient is no longer super-Eddington is plotted in red for each black hole mass in Figure \ref{fig:supereddtime}. The relation is not a simple power law, as the times for which $L_{\rm acc}$ is super-Eddington drops steeply for $M_{\mathrm{h}} \geq 5\times 10^6 M_{\odot}$ compared to the lowest three black hole masses.

\section{Discussion}
\label{sec:discussion}
The value of $q$ constrains the density and pressure profiles of the ZEBRA envelope, with larger $q$ implying a more spherical envelope. Our results from the numerically-simulated fallback rates therefore indicate that the accretion disk ``puffs up'' more at early times than the analytic, frozen-in model predicts (see Figure \ref{fig:q16}). 
This finding implies that, if the jet remains in pressure balance with the ZEBRA envelope (equivalently, if the ZEBRA is the source of the collimation for the outflow), the jet containing the released accretion energy is more highly collimated at early times.

\begin{figure}
\includegraphics[width=0.495\textwidth]{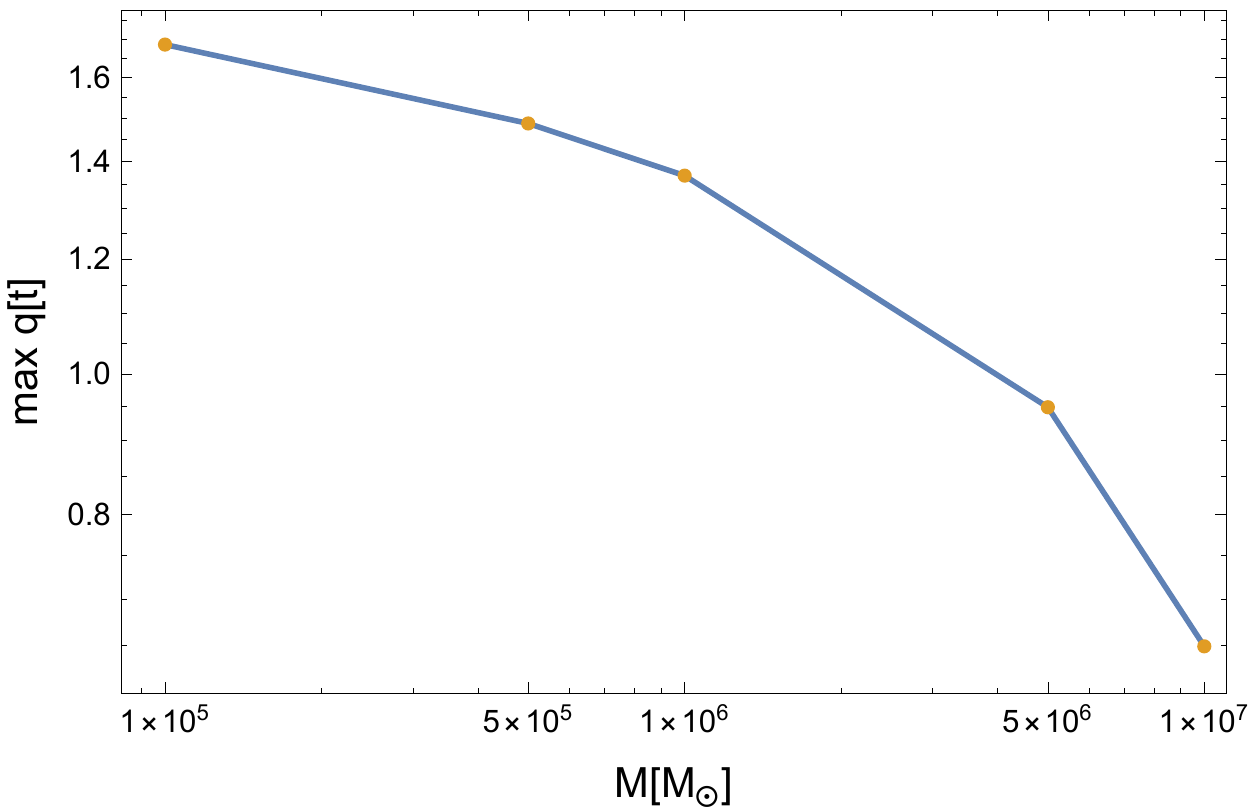}
\caption{Maximum $q$ as a function of black hole mass for 5 equally-spaced discrete values between $10^5$ and $10^7 M_{\odot}$. Line shown is not a best fit but merely for clarity of visualization.}
\label{fig:maxq}
\end{figure}

By plotting the maximum value of $q$ achieved for each black hole mass as in Figure \ref{fig:maxq}, we find that higher mass black holes attain lower maximum values of $q$ than a power-law would predict. Qualitatively, since $q$ is inversely proportional to angular momentum $\mathcal{L}^{5/6}$ and $\mathcal{L}$ is proportional to the mass of the black hole $M_{\rm h}^{2/3}$, $q$ is ultimately inversely proportional to $M_{\rm h}^{5/9}$. Thus higher-mass black holes attain lower values of $q$. However, the dependence of $\mathcal{L}$ on the fallback rate, which itself depends on the black hole mass,leads to an approximate relationship $\mathcal{L}\propto M_{\rm h}^{2/3} - M_{\rm h}^{1/3}$. This prevents a simple power law relationship between $q$ and $M_{\rm h}$.

Our results offer an upper limit on the maximum black hole mass for which super-Eddington accretion occurs in TDEs, corresponding to $10^7 M_{\odot}$ (assuming a Solar-like star that is accurately described by a $\gamma = 5/3$ polytrope). Importantly, this upper mass limit is derived from the numerically-obtained fallback rates, while the equivalent limit obtained from the frozen-in approximation would be significantly lower ($\sim few\times 10^6M_{\odot}$). 
We find that super-Eddington accretion occurs on the order of one year for a black hole of mass $5\times10^6 M_{\odot}$, and for approximately one month for a $10^7 M_{\odot}$ SMBH.

\begin{figure}
\includegraphics[width=0.495\textwidth]{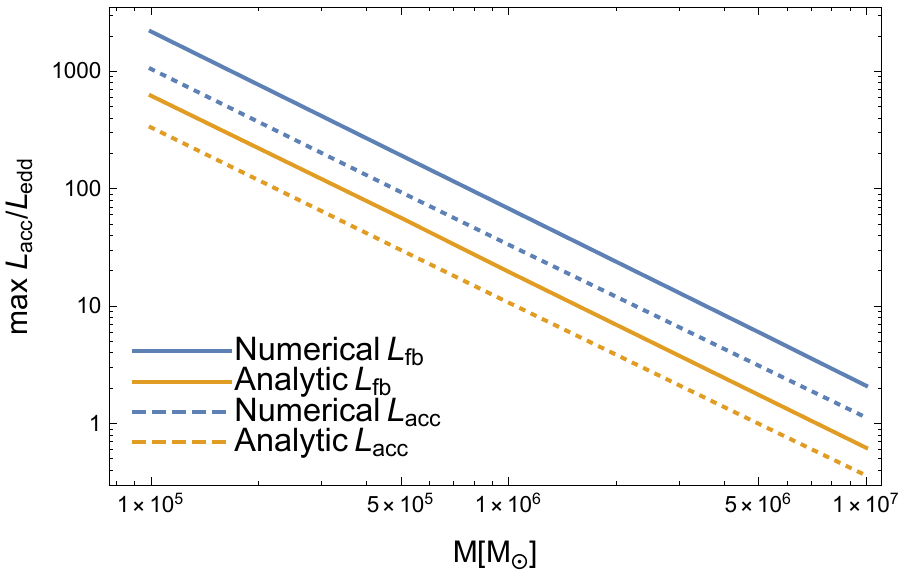}
\caption{Power law fits for maximum accretion luminosity (dotted) as a function of black hole mass for numerical (blue) and analytic (yellow) results. The maximum fallback luminosity (solid) from the numerical (blue) and analytic (yellow) prescriptions for the fallback rate are also shown with their power law fits. All are normalized by the Eddington luminosity with an assumed radiative efficiency of $\epsilon = 0.1$.}
\label{fig:maxmacc}
\end{figure}


The frozen-in approximation predicts that the fallback rate scales as (see Equation 34 of CB14)

\begin{equation}
\dot{M}_{\rm fb} = \frac{M_*}{t_{\rm r}}f(\tau), \label{mdotfb}
\end{equation} where 

\begin{equation}
t_{\rm r} = \left(\frac{R_*}{2}\right)^{3/2}\frac{2\pi M_{\rm h}}{M_* \sqrt{GM_{\rm h}}} \label{tr}
\end{equation}
is the return time of the most bound debris and $f$ is a numerically-obtained function of $\tau$ -- time normalized by $t_{\rm r}$ -- and the density profile of the stellar progenitor. Since Equation \eqref{mdotfb} only depends on the black hole mass through $t_{\rm r}$, we expect 

\begin{equation}
\frac{L_{\rm fb}}{L_{\rm Edd}} \propto \frac{\dot{M}_{\rm fb}}{M_{\rm h}} \sim M_{\rm h}^{-3/2}.
\end{equation} 
Thus, the frozen-in approximation predicts that the maximum fallback rate normalized by the Eddington luminosity follows a $M_{\rm h}^{-3/2}$ power law relationship.

Figure \ref{fig:maxmacc} depicts the maximum accretion luminosity normalized by the appropriate Eddington luminosity for each black hole. The numerical fallback results in a power law relationship between the accretion luminosity (normalized by the appropriate Eddington luminosity) and black hole mass: 

\begin{equation}
\frac{L_{\rm acc}}{L_{\rm Edd}} = 20.4 M_{\rm {h,6}}^{-1.46}, \label{maxaccnumerical}
\end{equation}
where $M_{\rm h,6}$ is the SMBH mass in units of $10^6 M_{\odot}$. The frozen-in approximation also leads to a power law for the accretion luminosity: 

\begin{equation}
\frac{L_{\rm acc}}{L_{\rm Edd}} = 6.3 M_{\rm h,6}^{-1.46}. \label{maxaccanalytic}
\end{equation}
Comparing these two expressions, we see that the maximum numerical accretion luminosity scales three times greater than the maximum analytic accretion luminosity.

It has been shown that self-gravity, which the impulse approximation neglects, can be important for modifying the structure of the tidally-disrupted debris stream and the fallback rate \citep{coughlin15,coughlin16b}. Because the fallback time scales with the square root of the SMBH mass (Equation \ref{tr}), one might suspect that deviations from the expected $L_{\rm fb}/L_{\rm Edd} \sim M_{\rm h}^{-3/2}$ scaling would be more pronounced for larger black hole masses (as self-gravity has more time to act in these cases), leading ultimately to a power law index that differs from $-3/2$. However, this is not the case: the power-law index remains the same for both the numerically-obtained and the analytic fallback rates, and is very nearly equal to $-3/2$. This finding suggests that self-gravity is most important for modifying the fallback rate at early times when the density of the stream is highest. We also propose that these early deformations induced by self-gravity, which cause the tidally-disrupted debris stream to generate ``shoulders'' in its density profile (see Figures 6 and 7 of \citealt{coughlin16}), are responsible for the factor of $\sim 3$ discrepancy between the analytical and numerical peak accretion rates.

In the context of observed tidal disruption events, the ZEBRA model of CB14 was tailored to explain jetted, super-Eddington TDEs. In the model, the luminosity produced within the accretion disk is necessarily exhausted anisotropically from the ZEBRA through bipolar jets owing to the supercritical nature of the accretion. In agreement with this condition, the three TDEs so far observed that display jetted activity -- \emph{Swift} J1644+57 (e.g., \citealt{bloom11}), \emph{Swift} 20158+05 \citep{cenko12}, and \emph{Swift} J1112 \citep{brown15,brown17} -- were likely accreting at a super-Eddington rate. In particular, J1644 and J1112 were associated with black holes of respective mass $M_{\rm h} \simeq 3\times 10^6M_{\odot}$ \citep{levan16} and $M_{\rm h} \gtrsim 5\times10^6 M_{\odot}$ \citep{brown15}, and though there are uncertainties related to the beaming factors of the jets (the isotropic luminosities were well above the Eddington limit for these SMBHs), the luminosities of those systems were likely super-Eddington by factors of at least 10-100. 
The mass of the SMBH powering J2058 could only be constrained to $M_{\rm h} \lesssim 8\times10^6$, but the isotropic luminosity of the system ($L_{\rm X} \simeq 3\times10^{47}$ erg s$^{-1}$) was still in excess of the Eddington limit of even the largest conceivable black hole mass. 
The approximate, $\propto t^{-5/3}$ decline of the lightcurves of each of the jetted TDEs also fits well with the expectations of the ZEBRA model. 
We therefore conclude that, overall, the ZEBRA model describes qualitatively well the observed X-ray lightcurves of the known jetted TDEs.

One of the other predictions of the ZEBRA model is that the time at which the accretion rate falls below Eddington, after which the jetted activity should cease, is a decreasing function of SMBH mass (cf.~Figure \ref{fig:supereddtime}). 
Using the above-quoted masses for J1644 ($10^6 M_{\odot}$), J2058 ($8\times 10^6M_{\odot}$), and J1112 ($5\times 10^6M_{\odot}$), the ZEBRA model predicts a shorter lifetime of a few months to three-quarters of a year for J1112 and J2058, and a longer lifetime of $\sim 1-2$ years for Swift J1644. These jetted-activity lifetimes are consistent with observations of Swift J1112, which exhibited a sharp decline in the X-ray luminosity after $\sim 40$ days; Swift J2058, whose X-ray luminosity was observed for timescales on the order of months; and Swift J1644, whose luminosity declined precipitously after $\sim 1.4$ years.

Finally, optical/UV observations of Swift J2058 during the jetted phase established a roughly constant effective temperature of $T\gtrsim 6\times 10^4 K$ (\citealt{cenko12}; J1644 displayed no optical/UV emission, presumably from dust extinction, and J1112 was only found archivally). From our Figure \ref{fig:rTplots}, the ZEBRA model predicts that a roughly constant temperature should be established during a super-Eddington TDE, and a value of $6\times 10^4$ K corresponds to a black hole of mass $\sim 5 \times 10^6M_{\odot}$. This value is consistent with the constraints on the black hole mass for Swift J2058 \citep{cenko12}.

Interestingly, observed optical and UV TDEs that lack hard X-ray and radio (and are therefore probably not jetted) also 
exhibit approximately constant effective temperatures around $\sim$ few$\times 10^4$ K (e.g., \citealt{gezari12,strubbe15,hung17}). 
In contrast, thin disk models of TDEs estimate an effective temperature of $T\simeq 10^5 K$ \citep{cannizzo90} -- an order of magnitude in excess of those observed. While these optical and UV TDEs are not jetted, the agreement between our predictions and the observations of the temperature evolution, while not conclusive, could imply that these optical/UV TDEs are also well-parameterized by a zero-Bernoulli condition.

\section{Summary and Conclusions}
\label{sec:summary}
Tidal disruption events, which occur when a star is destroyed by the tidal field of a supermassive black hole, can exhibit super-Eddington accretion as tidally disrupted debris falls back to the hole. During this super-Eddington phase, the debris disk formed from the TDE cannot cool efficiently, and the disk likely becomes highly inflated and weakly bound; outflows, either in the form of jets or winds, are also likely generated. \citet{coughlin14} (CB14 in this paper) used these notions to construct a self-similar, analytic model of the accretion disk structure formed from a TDE, denoting the structure a zero-Bernoulli accretion (ZEBRA) flow. Under this paradigm, accretion onto the SMBH is approximately super-Eddington, and the excess accretion energy that would otherwise unbind the disk is exhausted from the system through bipolar jets  launched along the rotational axis of the system. 

Our goal in this paper was to estimate the duration and magnitude of super-Eddington accretion in TDEs, and specifically to understand how the mass of the disrupting SMBH affects these estimates. To this end, we first ran a suite of numerical simulations of the tidal disruption of a Solar-like star, modeled as a $\gamma = 5/3$ polytrope, by a black hole with a mass in the range $10^5$ -- $10^7 M_{\odot}$. From these simulations we directly calculated the fallback rate -- the rate at which tidally-disrupted debris returns to pericenter following a TDE -- as a function of black hole mass (Section \ref{sec:simulations}). We then used the ZEBRA model of CB14 in conjunction with our numerically-obtained fallback rates, which ``feed'' the ZEBRA, to analyze the time-dependent structure of the accretion disk formed during the TDE and the accretion rate onto the SMBH (Section \ref{sec:accretion}). 

Our fallback rates were calculated from numerical simulations of TDEs, which include the effects of self-gravity and pressure on the long-term evolution of the debris stream. This direct calculation is more accurate than the impulse, or ``frozen-in,'' approximation used by CB14, which models the orbits of tidally-disrupted gas parcels as Keplerian following disruption. Comparing the two approaches, we found that the numerically-obtained, more realistic fallback rate predicts a peak accretion luminosity larger by nearly an order of magnitude than that of the impulse approximation. We also found that the ZEBRA envelope is characterized by a steeper radial density gradient and a more spherical gas distribution when the fallback rate is prescribed by the numerical simulations. 

The ZEBRA envelope becomes less spherical and achieves smaller peak accretion luminosities as the SMBH mass increases. Furthermore, our numerical fallback rates demonstrate that for SMBHs with mass in excess of $10^7 M_{\odot}$, the accretion luminosity is always sub-Eddington and therefore invalidates the ZEBRA prescription. In contrast, the fallback rate derived from the impulse approximation places this limiting SMBH mass at $\sim 5\times 10^6M_{\odot}$. The realistic fallback rates thus raise the upper limit on the maximum mass for super-Eddington accretion in tidal disruption events to values above prior extrema derived from the impulse approximation.

In this paper, the tidal disruption event was simulated up to obtaining the fallback rate, and we did not go on to numerically investigate the accretion disk formation. As a result, conclusions related to the properties of the accretion disk face limitations. In particular, the ZEBRA model assumes that the debris stream circularizes efficiently. Should the debris stream instead miss itself, for example due to Lens-Thirring precession, the fallback and accretion rates could differ appreciably \citep{guillochon15,hayasaki16}. 

Our models assumed that the star was Solar-like and possessed a $\gamma = 5/3$ polytropic density distribution. In reality, there is a range of stellar masses and properties intrinsic to each galaxy, and polytropes -- while a good first estimate -- are probably insufficient for capturing the intricacies of the density profiles of most stars. Therefore, a more detailed study characterizing the dependence of super-Eddington accretion on stellar properties is required, and our results here should be considered as a small subset of those data points. We leave such a detailed study to a future investigation.

\section*{Acknowledgments}
ERC acknowledges support from NASA through the Einstein Fellowship Program, grant PF6-170150. CN is supported by the Science and Technology Facilities Council (grant number ST/M005917/1). The Theoretical Astrophysics Group at the University of Leicester is supported by an STFC Consolidated Grant. This research used the Savio computational cluster resource provided by the Berkeley Research Computing program at the University of California, Berkeley (supported by the UC Berkeley Chancellor, Vice Chancellor for Research, and Chief Information Officer). We also thank the referee for useful comments and suggestions.

\bibliographystyle{mnras}
\bibliography{bib}

\bsp	
\label{lastpage}
\end{document}